\documentclass[preprint]{elsarticle}
\usepackage{amssymb}
\usepackage{amsfonts}
\usepackage{amsthm}
\usepackage{amsmath}
\usepackage{fancyhdr}
\usepackage{pstricks}
\usepackage{epsfig}
\usepackage{enumitem}
\usepackage{graphicx}
\usepackage {longtable}
\usepackage{xcolor}
\usepackage{tikz}
\usetikzlibrary{positioning}

\newcommand{\Cc}{\ensuremath{\mathcal  C}}

\newcommand{\Gc}{\ensuremath{\mathcal  G}}
\newcommand{\Hc}{\ensuremath{\mathcal  H}}

\newcommand{\Sc}{\ensuremath{\mathcal  S}}

\newcommand{\dr}{\ensuremath{\mathrm{d}}}

\newcommand{\bF}{\ensuremath{\mathbb{F}}}

\newcommand{\beg}{\begin{example}}
\newcommand{\eeg}{\end{example}}
\newcommand{\bit}{\begin{itemize}}
\newcommand{\eit}{\end{itemize}}
\newcommand{\bcor}{\begin{cor}}
\newcommand{\ecor}{\end{cor}}
\newcommand{\beq}{\begin{equation}}
\newcommand{\eeq}{\end{equation}}
\newcommand{\beqn}{\begin{equation*}}
\newcommand{\eeqn}{\end{equation*}}
\newcommand{\bea}{\begin{eqnarray}}
\newcommand{\eea}{\end{eqnarray}}
\newcommand{\bean}{\begin{eqnarray*}}
\newcommand{\eean}{\end{eqnarray*}}
\newcommand{\ben}{\begin{enumerate}}
\newcommand{\een}{\end{enumerate}}
\newcommand{\bdefn}{\begin{definition}}
\newcommand{\edefn}{\end{definition}}
\newcommand{\bnote}{\begin{note}}
\newcommand{\enote}{\end{note}}
\newcommand{\bprop}{\begin{proposition}}
\newcommand{\eprop}{\end{proposition}}
\newcommand{\blem}{\begin{Lemma}}
\newcommand{\elem}{\end{Lemma}}
\newcommand{\bthm}{\begin{theorem}}
\newcommand{\ethm}{\end{theorem}}
\newcommand{\pg}{\ensuremath{\mathrm{PG}(3,2)}}
\newcommand{\pgft}{\ensuremath{\mathrm{PG}(4,2)}}
\newcommand{\pgfq}{\ensuremath{\mathrm{PG}(4,q)}}
\newcommand{\pgfi}{\ensuremath{\mathrm{PG}(5,q)}}
\newcommand{\glfiq}{\ensuremath{\mathrm{GL}(5,q)}}

\newtheorem{proposition}{Proposition}[section]
\newtheorem{definition}{Definition}[section]
\newtheorem{theorem}{ Theorem}[section]
\newtheorem{Lemma}{{ Lemma}}[section]
\newtheorem{example}{Example}[section]
\newtheorem{Remark}{Remark}[section]

\begin{document}
	
	\begin{frontmatter}
	\title{Intersection Patterns in Optimal Binary $(5,3)$ Doubling Subspace Codes }
	\author[1]{Anirban Ghatak\corref{cor1}%
		\fnref{fn1}}
	\ead{ghatak.anirban@gmail.com}
	\author[2]{Sumanta Mukherjee\fnref{fn2}}
	\ead{sumanm03@in.ibm.com}
\cortext[cor1]{Corresponding author}
\address[1]{R.C. Bose Centre for Cryptology and Security, ISI, Kolkata, India}
\address[2]{IBM Research, Bangalore, Karnataka, India}

\begin{abstract}

	Subspace codes are collections of subspaces of a projective space such that any two subspaces satisfy a pairwise minimum distance criterion. There exists a significant body of research on subspace codes, especially with regard to their application for error and erasure correction in random networks. Recent results have shown that it is possible to construct optimal $ (5,3) $ subspace codes from pairs of partial spreads in the projective space $ \pgfq $ over the finite fields $ \bF_q $, termed doubling codes. In this context, we have utilized a complete classification of maximal partial line spreads in $ \pgft $ in literature to establish the types of the spreads in the doubling code instances obtained from two recent constructions of optimum $ (5,3)_q $ codes, restricted to $ \bF_2 $. Further we present a new characterization of a subclass of doubling codes based on the intersection patterns of key subspaces in the pair of constituent spreads. This characterization is a first step towards identifying all possible spread pairs that can yield optimal $ (5, 3)_2 $ doubling codes.

\end{abstract}
\end{frontmatter}

\section{Introduction}
The notion of subspace coding for errors and erasures in random networks was introduced in the papers of K\"{o}tter \emph{et al.} (\cite{KK, SKK}). There exists a rich body of literature dealing with the encoding and decoding of constant dimension codes, (for instance, references \cite{KoKu}-\cite{GT}) where all the codeword subspaces have the same dimension. In comparison, the more general problem of the construction of non-constant dimension subspace codes, often termed projective space codes, has not been addressed as thoroughly. An extensive list of open problems in this field and relevant references are to be found in Etzion's survey \cite{Etzion}. A motivation for investigating the construction of projective space codes is the intuition that a (size-)optimal subspace code is more likely to be obtained if the constraint of constant dimension codewords is relaxed. However, for several sets of parameters, there exist optimal constant dimension codes which achieve the tightest bounds on code size. Tables of existing constructions and bounds on the sizes of codes are available in \cite{HKKW2016} (with the link to a database). \\
Etzion and Vardy \cite{EV} proved a Gilbert-Varshamov-type lower bound and a linear programming upper bound on the size of projective space codes. In addition, they gave an example of a nearly optimal binary code of minimum \emph{subspace distance} $ 3 $ in a $ 5 $-dimensional ambient space, where the subspace distance between two subspaces $ U, V $ of an ambient projective space is defined in \cite{KK} as: $ {\dr}_s(U,V) :=\dim U+ \dim V - 2\dim (U \cap V) $.\\
An elegant ensemble of techniques for constructing subspace codes, which has resulted in some of the best known codes for certain parameters, is the so-called Ferrers diagram Rank-metric (FDRM) construction, cf. for instance, \cite{ES}, and more recently, \cite{EtZeh}, \cite{ZhanGe}. In \cite{KhK} the construction of constant dimension FDRM codes in \cite{ES} has been adapted for constructing projective space codes. Another approach ( for instance, in \cite{ES, ST, HKK}) involves \emph{puncturing} known constant dimension codes and adding suitable subspaces to increase the overall code-size.\\
An optimal binary $(5,3)$ projective space code was reported in \cite{AG1}, which was obtained by a strategy of minimal changes to the nearly optimal code given by Etzion and Vardy in \cite{EV}. Both the Etzion-Vardy (EV) code and the optimal variant can be described in terms of \emph{maximal partial spreads} of $2$-dimensional (vector) subspaces in a $5$-dimensional ambient (vector) space.  If we consider the associated projective geometry $ \pgft $, the $2$-subspaces and $3$-subspaces are the \emph{lines} and \emph{planes} of the geometry, respectively. In obtaining the optimal code in \cite{AG1}, results from the classification of maximal partial line spreads of $ \pgft $, as presented in \cite{RS} (cf. \cite{HKK, HKK14}), were used.\\
\emph{The Doubling Construction:} Honold, Kiermaier and Kurz \cite{HKK} have described a method for generating optimal $(5,3)$ codes over $\bF_q$, based on the \emph{point-hyperplane shortening} of a constant dimension \emph{lifted} rank-metric Gabidulin code \cite{Gab} augmented by additional subspaces. Over $\bF_2$, their construction, henceforth referred to as the HKK construction, results in optimal codes of size $18$. It is shown that the basic building blocks of these codes are again maximal partial line spreads of the ambient projective space. This motivated the so-called \emph{doubling construction}, also presented in \cite{HKK}. In $ \pgft $, the doubling construction aims to find a pair of maximal line spreads which yields an optimal code comprising the nine lines of one spread plus a set of nine planes. The planes are the subspace complements of the lines of the second line spread and are often referred to in this article as \emph{dual planes}, with a slight abuse of terminology.\\
A recent work by Cossidente, Pavese and Storme \cite{CPS18} (cf. \cite{copastorme}) provides another general construction for optimal $ (5,3)_q $ subspace codes. Their construction for even characteristic yields the following basic configuration (along with three other variants): a maximal partial spread of $ q^3 + 1 $ lines and an equal number of planes, which intersect mutually in exactly a point, such that no member of the line spread is contained in any of these planes. One notes that this is precisely the configuration of the doubling construction. So, while the scope and sophistication of the Cossidente-Pavese-Storme (CPS) construction are greater, the simplicity of the doubling construction is an advantage when considering small parameter cases, like, for instance, the $ (5,3)_2$ case. Moreover, one common thread emerges in both the HKK and CPS constructions - the regulus structure inherent in the constituent lines of the line spread and the complemented line spread of the set of planes. For instance, the CPS construction explicitly uses the reguli associated with a hyperbolic quadric in the even characteristic construction. In the HKK construction, both the spreads in a doubling code belong to the same type according to the regulus pattern classification (cf. Theorems \ref{th:hokklin} and \ref{th:hokkdu}), thus providing an interesting, if implicit, connection.\\

\noindent In this article we re-visit optimal binary $(5,3)$ subspace codes obtained by the doubling construction on two maximal line spreads in $ \pgft $. The incentive for considering optimal codes of precisely these parameters is the availability of a complete classification of maximal line spreads of $\pgft$ in \cite{RS}. We have sought answers to the following question:\\

\noindent \emph{What are the necessary conditions that are satisfied by two maximal line spreads in $ \pgft $ to yield an optimal $ (5,3)_2 $ code by the doubling construction?}\\

\noindent In this regard, our main contribution is the characterization of a large class of optimal $(5,3)_2$ doubling codes, based on the structure of constituent spreads (cf. Theorem \ref{th:intpat}). Our characterization of the optimal codes is based on the \emph{intersection patterns} involving the reguli of the two spreads. We have obtained all possible intersection patterns which can arise if both the constituent spreads of the optimal code are of \emph{type X} (cf. \cite{RS}).\\
As a secondary contribution we have started by proving that, over $ \pgft $, the HKK construction involves a pair of spreads precisely of type X, while the CPS construction mostly involves spreads of type I$ \Delta $ (cf. Sections \ref{sec:baggu}, \ref{sec:sprdtype}). We have further obtained the possible intersection patterns of the HKK construction as a direct application of our classification results for type X doubling codes.\\
\emph{Organization:} We begin the next section by stating some relevant results regarding the classification of line spreads in $ \pgft $. A brief description of the HKK construction, along with the doubling concept, is followed by an outline of the CPS construction. Section \ref{sec:sprdtype} establishes the types of the spreads involved when both the HKK and CPS constructions result in doubling $ (5,3)_2 $ optimal codes. In Section \ref{sec:intpat}, we present a complete description of all possible intersection patterns among the line spreads of type X, when the line spreads produce an optimal $ (5,3)_2 $ code by doubling. The possible patterns put forth in the pattern classification result are illustrated with examples in the next section (Section \ref{sec:eggu}). Further, as an application, the results in Section \ref{sec:intpat} are applied to identify the possible intersection patterns in the HKK construction in Section \ref{sec:hokkpat}. We conclude with a brief discussion of the results and future research.
\section{Background and Preliminaries}\label{sec:baggu}
In this section we first state some definitions and results regarding the classification of maximal line spreads in $\pgft$ ( cf. \cite{HKK, HKK14, RS}). Next we briefly outline the details of the HKK construction and in particular, the so-called doubling construction (\cite{HKK}) of optimal codes in $ \pgft $. We conclude the section with a brief outline of the CPS construction.
\subsection{Maximal Line Spreads and their Classification in $ \pgft $ }
\begin{figure}
	\begin{center}
		\includegraphics[width= 4.2 in]{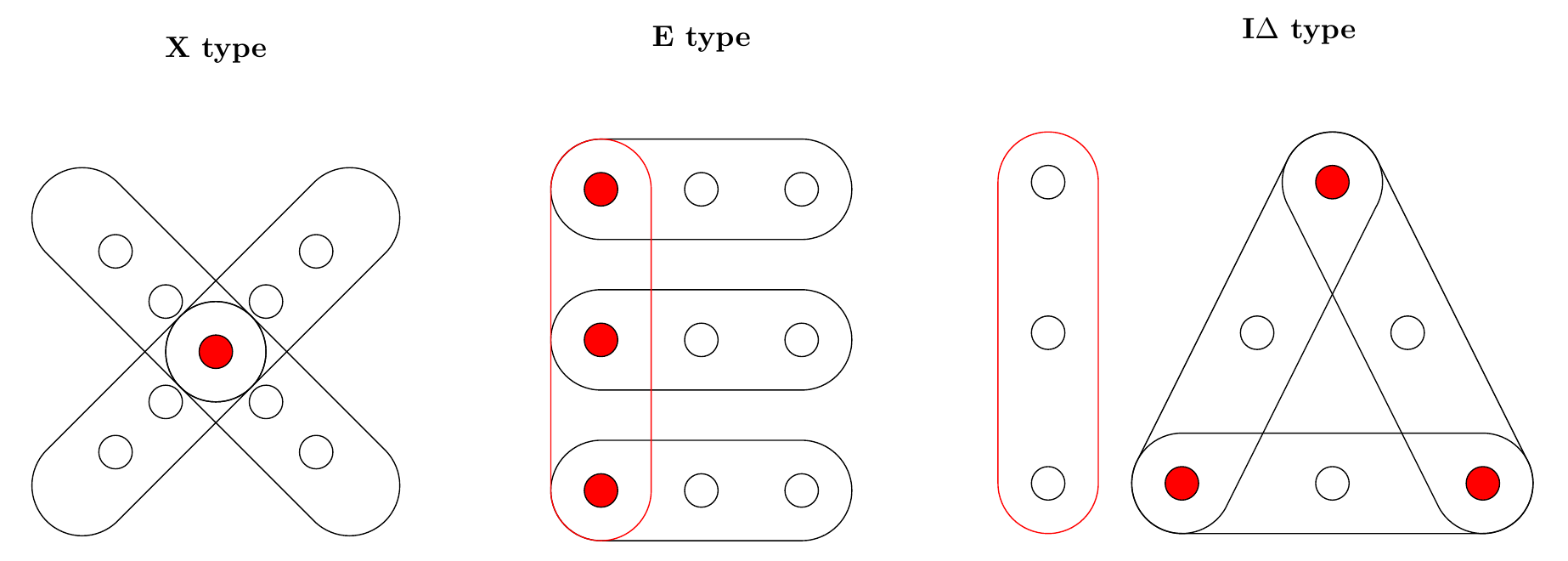}\\
		\caption{Types of Maximal Line Spreads of $\pgft$ }\label{fig:type}
	\end{center}
\end{figure}

We begin with a brief outline of the results on maximal line spreads in projective spaces.
\bdefn 
A \emph{regulus} in $\pgft$ is a set of three mutually non-intersecting lines which together span a hyperplane of $\pgft$, identified with $ \pg $, rather than the entire space. 
\edefn
The $9$ lines of a maximal line spread in $\pgft$ can be grouped into $4$ distinct reguli (cf. \cite{GSS}). Based on the distribution pattern of the lines among the $4$ reguli all possible maximal partial $2$-spreads in $\pgft$ have been shown (\cite{RS, GSS}) to belong to one of the following three types.
\begin{enumerate}
	\item A \textbf{type X} spread contains a particular line which belongs to all 4 reguli, the remaining lines belong to exactly one regulus each.
	\item Three lines form a distinguished regulus, which set-wise intersects each of the other three reguli in exactly one distinct line; the remaining six lines are disjointly shared among the three reguli. Such a spread has \textbf{type E}.
	\item In a spread of \textbf{type I$\Delta$}, three lines form one \emph{distinguished} regulus; three other lines are each shared by two reguli and the last triple of lines belong to one regulus each, the last six lines thus forming a $ \Delta $.
\end{enumerate}
In Figure \ref{fig:type}, the regulus patterns among the lines of the above three spread types are depicted. In all the diagrams, the lines are represented by circles and the triples forming each regulus are outlined. Those lines which are shared among multiple reguli are shaded - for instance, the ``ninth" line in a type X spread is common to all the reguli. Among the four reguli belonging to each spread, there exists a distinguished regulus which may be used to construct the so-called \emph{opposite} spread. In the patterns corresponding to the spread types E and I$\Delta$, the special regulus is readily identified. However, in the type X pattern, the identification of the special regulus requires some effort \cite{RS, GSS}.\\
As stated earlier, each regulus in $\pgft$ consists of three pairwise skew lines, which span a $\pg$. The $9$ points in the three lines of a regulus $ \rho $ can be set down in a $ 3 \times 3$ array, with each row corresponding to the points of a line. Choosing the points of this array column-wise returns another set of three lines which form the \emph{opposite regulus}, denoted $ \rho^o $. The six lines of $ \rho $ and $ \rho^o $ constitute the generators of a \emph{hyperbolic quadric} in the ambient $ \pg $. Replacing the distinguished regulus in a type X spread with its opposite regulus returns a type E spread and vice versa. Performing a similar operation on a type I$\Delta$ spread returns another type I$\Delta$ spread.
\subsection{Optimal $(5,3)$ Codes and the HKK Construction }
A $(5,3)_q$ subspace code is a subset of $\pgfq$ which can include subspaces of all possible dimensions with pairwise minimum subspace distance $3$. In \cite{HKK}, all possible dimension distributions as well as the achievable size of optimal $(5,3)_q$ subspace codes are specified. The size of such a code is $2q^3 + 2$ and a possible realization is the union of $q^3 + 1$ lines ($2$-subspaces) and an equal number of planes ($3$-subspaces). For instance, such an optimal code will consist of $9$ lines and $9 $ planes over $ \bF_2 $. Other realizations involve replacing one line by a point and/or replacing one plane by a hyperplane  in the above.\\
\emph{The HKK Construction:} The $(6, 3, 2)$ lifted Gabidulin code $\Gc$ is a set of $3$-subspaces of a $6$-dimensional vector space ($\pgfi$), obtained by the K\"{o}tter-Kschischang lifting \cite{KK} of a rank-metric Gabidulin code of $3 \times 3$ matrices with rank distance ${\delta}_r = 2$. The minimum subspace distance of $\Gc$ is ${\dr}_s = 2{\delta}_r = 4$. The \emph{point-hyperplane shortening} is performed on the set of subspaces of $\Gc$, plus two additional $3$-subspaces (planes of $\pgfi$) at subspace distance $4$. The shortening point is outside the special plane $\Sc$ of $\pgfi$ disjoint from all codewords of $\Gc$ (cf. Section $2.4$, \cite{HKK}) and the shortening hyperplane contains neither the special plane nor the point. The simultaneous shortening of $\Gc$ with respect to such a point $P$ and a hyperplane $H = \pgfq$ is described as:
\beqn
\Gc \lvert^{P}_{H} = \{ X \in \Cc \, \lvert \, X \subset H \}\cup \{Y\cap H, Y \in \Cc \, \lvert \, P \in Y \}
\eeqn 
The lifted code provides sets of $2$- and $3$-dimensional subspaces, each of size $q^3$, via shortening; one of the additional pair of $3$-subspaces furnishes a $2$-subspace via shortening, the other is included as a codeword. Thus we have a $(5, 3)_q$ code of size $2q^3 + 2$ with dimension distribution $(2,3)$.\\
\emph{The Doubling Construction:} Restricted to $\pgft$, the majority of the codes obtained via the HKK construction can be described as the union: $S1 \cup (S2)^{\perp}$, where $S1, S2$ are maximal line spreads in $\pgft$. The term $(S2)^{\perp}$ is the dual spread of $S2$, consisting of element-wise dual planes corresponding to the lines of $S2$. This is known as the \emph{doubling construction}.
\subsection{The CPS construction for even characteristic:}
Cossidente \emph{et al.} \cite{CPS18} have detailed similar but separate constructions of optimal subspace codes in $ \pgfq $ when $ q $ has characteristic odd and even, respectively. We next briefly outline the even characteristic construction which is relevant for optimal doubling codes in $ \pgft $.\\
The CPS construction yields four possible dimension distributions of subspaces constituting the optimal codes of which the Type IV (cf. Section $ 1.1 $, \cite{CPS18}) codes correspond to doubling. The steps of the construction are outlined as follows.
\begin{itemize}
	\item Define plane $ \pi $ and a line $ l \subset \pi $, $ \Hc $ a hyperbolic quadric which includes the regulus $ R_1 $ containing $ l $. Let $ R_2 $ be the opposite regulus of $ R_1 $.
	\item Define a subgroup $ G $ of the general linear group $ \glfiq $ as follows.
	\[G = \{ M_{a,b,c,d} \, \lvert \, a \in \bF_{q}^{\ast}, b,c,d \in \bF_q, c^2 + cd + \alpha d^2 = 1 \}\]
	where $ \alpha \in \bF_q $ is such that $ f(X) = X^2 + X + \alpha $ is irreducible over $ \bF_q $ and 
	\[M_{a,b,c,d} = \begin{bmatrix}
	1 & 0 & 0 & 0 & 0\\
	0 & ac & \alpha ad & ac & \alpha bd\\
	0 & ad & a(c+d) & bd & b(c+d)\\
	0 & 0 & 0 & a^{-1}c & \alpha a^{-1}d\\
	0 & 0 & 0 & a^{-1}d & a^{-1}(c+d)
	
	\end{bmatrix} \]
	\item Denote $ P_1 $ to be a \emph{good plane orbit} of $ G $, which consists of $ q^3 -q $ planes which pairwise intersect in a point. Let $ P_2 $ be a set of $ q +1 $ planes generated by the lines of the regulus $ R_1 $ and a point $ N $.
	\item Denote $ L_1 $ to be a \emph{good line orbit} of $ G $, which consists of $ q^3 -q $ mutually non-intersecting lines. Moreover, no line of $ L_1 $ is contained in any of the planes of $ P_1 \cup P_2 $. Likewise define $ L_2 $ to be the set of $ q + 1 $ mutually disjoint lines that constitute the regulus $ R_2 $.
\end{itemize}
The above construction yields a set of $ (q^3 -q) + (q+1) = q^3 +1 $ plane codewords, all of which pairwise intersect in a point. In addition it provides an equal number of mutually disjoint lines, none of which is contained in any of the plane codewords. Hence this configuration is indeed an instance of doubling, a maximal line spread and a set of planes which are complements of the lines of another maximal spread. Two variations of the above construction are also indicated in Remark $ 3.10 $ of \cite{CPS18}, for fixed choices of the so-called `good' plane and line orbits as follows.
\begin{enumerate}
	\item Given $ P_1, P_2, L_1, R_1 $ defined in the construction described above, consider the set $ P_3 $ of $ q +1 $ planes formed by the lines of $ R_2 = R_1^{op} $ and the point $ N $. Then $ P_1 \cup P_3 \cup L_1 \cup R_1 $ is another optimal doubling code where the `roles' of the reguli $ R_1 $ and $ R_2 $ in constituting the planes and lines outside the good plane and line orbits have been interchanged.
	\item The MLS of line codewords is described as $ L_1 \cup R_2 $, as in the original construction, but the set of plane codewords outside the good plane orbit is chosen differently. To elaborate, let $ \Sigma = \mathrm{PG}(3,q)$ be the hyperplane corresponding to the hyperbolic quadric $ \Hc $, associated with the regulus $ R_1 $. Replace a plane $ \pi \in P_2$ with another plane $ \pi' $, which contains the line $ l $ (cf. CPS construction details as above) but is itself not contained in $ \Sigma $, to complete the set of $ q+1 $ codeword planes.
\end{enumerate} 

\section{The Types of Spreads in HKK and CPS Doubling Constructions}\label{sec:sprdtype}
In this section, based on the existing results discussed in Section \ref{sec:baggu}, we identify the types of spreads involved when the HKK and the CPS constructions of optimal $ (5,3)_2 $ codes yield instances of doubling.
\subsection{The Type of Spreads in HKK Doubling}
 First we decide the types of the line spread $ S1 $, forming the set of line codewords, and the line spread $ S2 $ formed by the complements of the plane codewords in the HKK construction.
\subsubsection{The Spread Type of the Lines}
The maximal line spread (MLS) $S1$ that yields the lines in the HKK construction is characterized over $\bF_q$ in \cite{HKK} as follows: the $q^k$ uncovered points (holes) of $S1$ form the set-wise complement of a $k$-subspace $X_0 \in S$ in a $(k+1)$-subspace $Y_0$. The subspace $X_0$ is termed the \emph{moving subspace} of $S$, as it can be replaced by any $k$-subspace of $Y_0$ to form another MLS. To identify the type of MLS obtained as the line spread in the construction, when restricted to $\bF_2$, we first state the following result. 
\bthm [\cite{RS}, Theorem 3.2]\label{thm:Xcls}
(i) The partial spread $S_8 = \{ \lambda_1, \lambda_2,  \cdots, \lambda_8 \}$ which arises from any partition of $\pgft$ of the form $\lambda_1\cup \lambda_2 \cup \cdots\cup  \lambda_8 \cup \alpha_9  $, $\alpha_9$ a plane, is regulus-free.\\
(ii) If, in the above, $\lambda_9 $ is any line of the plane $\alpha_9$, then $S_9 = \{ \lambda_1, \lambda_2,  \cdots, \lambda_8, \lambda_9 \} $ is a partial spread in $\pgft$ of type X. $\blacksquare$
\ethm
Based on the description in \cite{HKK} and the above theorem, we have the following
\bthm\label{th:hokklin}
The MLS $ S1 $ in the HKK construction (over $ \bF_2 $) is of type X.
\ethm
\begin{proof} 
From the description in the HKK construction, the ``moving line " $X_0$ is chosen from the plane $Y_0$, identified with $E_1 = (\Sc + P)\cap H$ in \cite{HKK}, which contains the holes of the spread (cf. Sections $3.2$ and $3.3$, \cite{HKK}). Therefore we have: $Y_0 = X_0 \cup \{\textrm{holes of} \,\, S_9\}$. The partial spread $S_8$ of $8$ lines is obtained by shortening those codewords of $\Gc$, the lifted Gabidulin code, which contain the point $P$. Hence the shortened lines forming $S_8$ are all disjoint from $Y_0$. So the partition $S_8 \cup Y_0 = \pgft$ is of the type given in Theorem \ref{thm:Xcls} (i), and the line $X_0 \subset Y_0$ completes the MLS $S_9$ when added to $S_8$. Hence, the $S_8$ of lines is regulus-free, and the resulting MLS is of type X.
\end{proof}
\subsubsection{The Type of the Complemented Spread}
For identifying the type of the complemented or \emph{dual} spread $ S2 $, we refer to Remark $4$ of Section $3.3$ in \cite{HKK}. It is stated that the spread obtained in the dual form has the same type as the uncomplemented spread. We give a proof of this assertion by analyzing the construction of the dual spread in view of the results in \cite{RS}.\\
The set of dual planes obtained in the HKK construction comprises of $8$ codewords of the lifted $(6,3,2)$ Gabidulin code $\Gc$ and another additional plane, denoted $E' = E_2 $ in \cite{HKK}. The set of $8$ codewords are disjoint from a special plane $\Sc$ in $PG(5,2)$ which meets the shortening hyperplane $H = \pgft $ in a line. The ninth plane $B_9 = E_2$ meets the plane $\Sc$ in a line, denoted $L_2$ in \cite{HKK}.\\
In the description of a type X MLS in \cite{HKK}, the common line $ \lambda_9 $ is termed as the `moving' line, contained in the plane $ \alpha_9 $, which is the complement of the union of the remaining eight lines in $ \pgft $. Naturally, the dual configuration is a `moving' plane that contains a special line, which is identified as $ L_2$. We first fix $B_9$ as the moving plane, i.e. $ B_9 = {\lambda_9}^{\perp} $, the dual of the moving line of some type X spread. The special line $ L_2 \subset B_9$ is identified with the meet of the solids $ S_i $, which are the complements of the holes $ h_i $ of the line spread. Then we establish that the remaining planes $ B_i, \, i = 1, \cdots, 8 $, can be identified as the duals of lines $ \lambda_i, \, i = 1, \cdots, 8 $, of the spread, in some order.
\bthm\label{th:hokkdu}
The MLS $ S2 $, formed by the complements of the plane codewords obtained by the HKK construction (over $ \bF_2 $), is of type X.
\ethm
\begin{proof}
	Recall that a maximal partial line spread of type X arises out of a partition of $\pgft$ of the form \cite{RS}: 
	\beqn \label{eq:xsprd} 
	S= \lambda_1\cup \lambda_2 \cup \cdots\cup  \lambda_8 \cup \alpha_9 
	\eeqn
	where $\alpha_9$ is a plane. The ninth line of the spread is any line $\lambda_9 \subset \alpha_9$ and so, the $4$ holes of the spread are given by $\{h_1, h_2, h_3, h_4\} = \alpha_9 \setminus \lambda_9$. The duals of the holes $h_i, \, i = 1,\cdots , 4$, are hyperplanes in $\pgft$, i.e. solids denoted by $S_i,  \, i = 1,\cdots , 4$. Therefore, we have:
	\beqn
	\left( \bigcup_{i=1}^{4} h_i \right) ^{\perp} =  \bigcap_{i=1}^{4} {\left( h_i \right) ^{\perp}} = \bigcap_{i=1}^{4} S_i
	\eeqn
	If the ninth plane $B_9$ is identified as ${\lambda_9}^{\perp}$, and $ L_2 \subset B_9 $ is as described above, it remains to identify the planes $B_i $ as the dual planes $ {\lambda_i}^{\perp}, \, i = 1, \cdots, 8$. To justify this choice, we claim it is enough to prove that all the plane codewords $ B_i,\, i= 1, \cdots, 8 $, are disjoint from $ L_2 $.\\
	To that end we have that any three of the holes $\{h_1, h_2, h_3, h_4\}$ are linearly independent and the union of the holes with any $\lambda_i,\, i = 1,\cdots, 8$, disjoint from $\alpha_9$, spans the entire ambient space. So any dual plane $ \lambda_{i}^{\perp} , \, i =1,\cdots, 8$, is disjoint from the line $ L_2 $ since
	\beqn 
	\pgft = \left( \bigcup_{i=1}^{4} h_i \right)\cup \lambda_{i} = \left(\left( \bigcup_{i=1}^{4} h_i \right) ^{\perp} \cap \lambda_{i}^{\perp} \right)^{\perp} = \left(L_2 \cap\lambda_{i}^{\perp} \right)^{\perp}.
	\eeqn	
	But, in the HKK construction, the plane codewords $B_i,\, i = 1,\cdots,8$, are all disjoint from $L_2 \subset \Sc$, the special plane disjoint from the codewords of the initial Gabidulin code $\Gc$. So, after possible re-ordering, $ B_i = \lambda_{i}^{\perp} $ for $i =1,\cdots, 8$, and the assertion is proved.\\
	Hence the set of planes obtained by the HKK construction of $(5,3)_2$ codes is the dual of a line spread of type X. 
\end{proof}
The upper bound on the size of an optimal $(5,3)_2$ code was proven by Etzion and Vardy to be $18$ (\cite{EV}); their proof also details the possible dimension distributions of such optimal codes. Out of these possibilities, Honold \emph{et al.} (\cite{HKK}) have found that the most numerous are the doubling codes constituted of nine lines and nine planes. We denote the lines of such an optimal code, which form a maximal line spread $S1$, as $a_1, a_2, \cdots, a_9$. The planes $B_1, B_2, \cdots, B_9$ are the complements of the lines of a second MLS $S2$. If either of the spreads is of type X, the special line, common to all the reguli, is chosen to be the ninth line in each case.\\
An analysis of optimal codes of this form, henceforth referred to as \emph{type X codes}, with respect to the regulus structure of the constituent spreads, is presented in Section \ref{sec:intpat}. 
\subsection{The Type of Spreads in CPS Doubling}
In this section we establish the types of the line spread formed by the set of line codewords, and that formed by the complements of the plane codewords in the basic Type IV CPS construction for $ (5,3)_2 $ optimal codes. To this end we first briefly state a few relevant results from \cite{GSS}.\\
From \cite{GSS} (Section $ 5.1 $), we know that there are three types of extensions of a partial spread of size $ 6 $, denoted $ \Sc_6 $, to an MLS $ \Sc_9 $ of size $ 9 $, depending on the regulus composition of the underlying $ \Sc_6 $. All three configurations of the $ \Sc_6 $, in turn, arise from a `projectively unique' $ \Sc_5(L) $, with five lines forming two reguli having a common line, denoted $ \lambda_1 $ ( cf. Section $ 4 $, \cite{GSS}). The regulus composition of the three types of $ \Sc_9 $ in terms of the underlying $ \Sc_6 = \Sc_5(L) \cup \{\lambda_6\} $ is described as follows.
\begin{enumerate}[label={(\roman*)}]
	\item Two extensions of an $ \Sc_6 $, containing exactly two reguli, to either an $ \Sc_9 (E) $ or an $ \Sc_9 (X) $:
	\begin{itemize}[label={--}]
		\item Among the complementary triple of lines, denoted $ \{\lambda_7, \lambda_8, \lambda_9\} $, say $ \lambda_7 $ completes a regulus with $ \lambda_6 $ and a line of $ \Sc_5(L) $, while $ \lambda_8, \lambda_9 $ complete a regulus with another line of $ \Sc_5(L) $, forming $ \Sc_9(E) $.
		\item The line tuples $ \{\lambda_6,\lambda_7\} $ and $ \{\lambda_8,\lambda_9\} $ form two reguli with the common line $ \lambda_1 $ of the $ \Sc_5 $, thereby forming $ \Sc_9(X) $.
	\end{itemize} 
	
	\item An extension of the $ \Sc_5(L)  $ to an $ \Sc_6 (\Delta) $, containing three reguli, is possible if $ \lambda_6 $ forms another regulus with one line each from the two reguli of $ \Sc_5$. This implies that $ \lambda_7, \lambda_8, \lambda_9 $ form the fourth regulus of $ \Sc_9(\mathrm{I}\Delta) $. 
\end{enumerate} 
Hence for $ \Sc_9(E) $ and $ \Sc_9 (X) $, the set of three lines complementing the underlying $ \Sc_6 $ do not form a regulus by themselves. Instead, these lines complete two reguli in conjunction with the lines of the $ \Sc_6 $. Further, in the composition of $ \Sc_9(E) $ and $ \Sc_9(\mathrm{I}\Delta) $, none of the lines outside the underlying $ \Sc_5 $ form a regulus with $ \lambda_1 $.
\subsubsection{The Spread Type of the Lines}
From the discussion in Section \ref{sec:baggu}, the basic construction for even characteristic yields $ 9 $ lines of an optimal $ (5,3)_2 $ code as follows. A so-called good line orbit furnishes $ 2^3 -2 = 6 $ pairwise disjoint lines; adding three more lines constituting a regulus $ R_1 $ or its opposite, completes the MLS of size $ 9 $. In the light of the preceding discussion we have the following
\bprop \label{prop:cpstyp1} 
The line spread in a binary CPS doubling code has type $ \mathrm{I}\Delta $.
\eprop
\begin{proof}
	The three lines of the $ \Sc_3 $, extending the $ \Sc_6 $ formed by the good line orbit, constitute a regulus ($ R_1 $ or $ R_1^{op} $). Hence the spread $ \Sc_6 $ necessarily contains the remaining $ 3 $ reguli of the MLS $ \Sc_9 = \Sc_6 \cup \Sc_3 $, which means it is an $ \Sc_6 (\Delta) $. The proposition follows.
\end{proof}

\subsubsection{The Type of the Complemented Spread}
Next we first identify the type of the spread formed by the line complements of the set of $ 9 $ planes produced by the basic Type IV construction.\\
It follows from (cf. Remark $ 3.7 $ of) \cite{CPS18} that within the plane codeword set, the $ 3 $ planes outside the good plane orbit are constructed from $ 3 $ pairwise disjoint lines, which form a regulus of $ \pgft $, and a specific point denoted $ N $. We know that the line complements of the plane codeword set are pairwise non-intersecting and hence, form an MLS $ \Sc_9 $. As in the previous subsection, we can identify this MLS as type $ \mathrm{I}\Delta $ if any of the following equivalent conditions hold.
\begin{enumerate}[label={(\roman*)}]
	\item The line complements of the set of $ 6 $ planes of the good plane orbit form an $ \Sc_6(\Delta) $ containing $ 3 $ reguli.
	\item The line complements of the remaining $ 3 $ planes form a regulus.
\end{enumerate}
To this end we have the following 
\bprop\label{prop:cpstyp2}
The line complements of the $ 3 $ plane codewords outside the good plane orbit in the basic Type IV CPS construction over $ \bF_2 $ constitute a regulus in $ \pgft $.
\eprop
\begin{proof}
	Let $ \lambda_7, \lambda_8, \lambda_9 $ be the line complements of the plane codewords $ \pi_7, \pi_8, \pi_9 $, respectively. As the planes pairwise intersect in the point $ N \in \pgft$, it follows that the lines are pairwise disjoint. Moreover, we have $ \{N\} = \pi_7 \cap \pi_8 \subseteq \pi_9$, or, equivalently, $ \pi_7^{\perp} \cup \pi_8^{\perp} \supseteq \pi_9^{\perp} $. This in turn implies that $ \lambda_9\subseteq \langle \lambda_7, \lambda_8 \rangle $, i.e. $ \lambda_7, \lambda_8, \lambda_9 $ constitute a regulus of $ \pgft $.
\end{proof}
Using Proposition \ref{prop:cpstyp1} and Proposition \ref{prop:cpstyp2}, we state the types of the spreads in the doubling description of the basic Type IV $ (5,3)_2 $ codes in the following
\bthm \label{thm:cpstyp}
The basic Type IV CPS construction for optimal $ (5,3)_2 $ codes is an instance of doubling with both the sets of line codewords and the line complements of the plane codewords identified as MLS's of type $ \mathrm{I}\Delta $.
\ethm 
\begin{Remark}
	It is readily seen that the first variant of the basic Type IV CPS construction, as presented in Section \ref{sec:baggu}, yields the same spread types when viewed as a doubling code. However, in the second variant, the spread type of the set of line complements may no longer be $ \mathrm{I}\Delta $. This is because replacing one of the plane codewords from the triple of planes intersecting in a common point may result in the line complements no longer forming a regulus by themselves. But the spread of line codewords is still of type $ \mathrm{I}\Delta $.
\end{Remark}
\section{Intersection Patterns in Type X Optimal $(5,3)_2$ Codes}\label{sec:intpat}
In this section we consider the intersection patterns of optimal $(5,3)_2$ codes, each of which can be expressed as the doubling $S1 \cup (S2)^{\perp}$ of two spreads of type X. Specifically we look at the intersection patterns of a dual plane $B$, which is the complement of a line of the second spread $S2$, with the reguli of the first spread $S1$. In the first subsection, we set down all feasible intersection patterns based on the properties of a type X spread. In the next subsection we eliminate some of the possible patterns using further restrictions to ensure the optimality of the resulting code. Finally we state our main result (Theorem \ref{th:intpat}) in the form of a complete classification of optimal $(5,3)_2$ codes of type X in terms of these intersection patterns.
\subsection{Feasible Intersection Patterns with a Dual Codeword Plane}
Since $S1$ is a spread of type X, the intersection pattern of the plane $B$, or any codeword plane for that matter, depends on the intersection of the said plane with the common line $a_9$ of $S1$. We show in the following lemma how the number of holes of $S1$ contained in the codeword plane characterizes the intersection pattern of the plane with the lines of $S1$. 
\blem \label{lemhole}
In the doubling construction of $(5,3)_2$ codes, the number of holes of the line spread $S1$ contained in a codeword plane $B$ of the complemented spread $S2$, is given as follows.
\begin{enumerate}
\item $B$ is disjoint from $a_9$ and contains only $1$ hole;
\item $B$ intersects with $a_9$ and either contains $2$ holes or none at all.
\end{enumerate}
\elem
\begin{proof}
For a minimum subspace distance $ 3 $, a codeword plane can either be disjoint from the line $a_9$ or intersect in a point.
\begin{enumerate} 
	\setlength\itemsep{1.2 mm}
\item $ B \cap a_9 = \phi $:\\
\noindent  The holes of $ S1 $, say $h_i, i = 1,\cdots, 4$, form the set complement $\alpha_9 \setminus a_9$, and hence, any three of them are linearly independent. As $B , \alpha_9$ are distinct planes in $\pgft$, $B \cap \alpha_9 $ is either a line or a point. If $B$ is disjoint from $ a_9 \subset \alpha_9 $, it contains exactly a point of $ \alpha_9 \setminus a_9 $, which is a hole of $S1$.
\item $B \cap a_9 = x$, a point:\\
\noindent Suppose, in this case, $B$ also contains a hole $h_1 \in \{ \alpha_9 \setminus a_9 \}$. Then $x + h_1 \in B$ is hole of $S1$. As stated in the previous case, $B$ cannot contain three or more holes. Hence $B$ either contains two holes of $S1$, such that they form a line with the point $B \cap a_9 = x$, or none at all.
\end{enumerate}
\end{proof}
Now we describe the feasible intersection patterns for a codeword plane $B$ with the four reguli of $S1$. An intersection pattern is described in the form of a tuple $ (i,j,k,l) $, where the entries of the tuples belong to the set $ \{0,1,2,3\} $. A single entry $ j \in \{1,2,3\}$ among the tuple indicates that $ B $ intersects with $ j $ lines of some regulus $ r $;  an entry $ 0 $ means that $ B $ is \emph{disjoint} from all three lines of a regulus. For instance, the tuple $ (3,\, 3, \, 1, \, 0) $ indicates that the codeword plane $ B $ intersects with all $ 3 $ lines of two reguli, with only a single line of another regulus and is disjoint from all the lines of the remaining regulus.
\subsubsection{ Disjoint Intersection Patterns}
In the first case of Lemma \ref{lemhole}, $B$ is disjoint from $a_9$ and contains $1$ hole. Hence the remaining six points of $B$ are distributed among six lines of the spread $S1$, excluding $a_9$. Therefore, we have the following feasible patterns:\\

\textbf{1.} $ (2,\, 2, \, 2, \, 0) $; \hspace{4.8 mm} \textbf{2.} $ (2, \, 2, \, 1, \, 1) $.\\

The above patterns are depicted as follows. The $4$ rows denote lines in the $4$ reguli of $S1$, bullets denote the lines intersecting with $B$ and circles denote the lines disjoint from $B$. Since $S1$ is of type X, we can assign the third (rightmost) column in the array to the common line $a_9$. Hence for a disjoint intersection pattern, all the entries in the third column are represented by circles.\\
\begin{center}
Pattern \textbf{1.} $ \,\,\,
\begin{matrix}
\bullet &\bullet &\circ\\ \bullet &\bullet &\circ\\ \bullet &\bullet &\circ \\ \circ &\circ &\circ
\end{matrix} $ $ \,\,\,\,\,$ Pattern \textbf{2.} $ \,\,\, \begin{matrix}
\bullet &\bullet &\circ\\ \bullet &\bullet &\circ\\ \bullet  &\circ &\circ \\ \bullet &\circ &\circ
\end{matrix}
$\\
\end{center}
\subsubsection{ Common Point Intersection Patterns}
In the second case of Lemma \ref{lemhole}, $B$ intersects $a_9$ in a point, denoted $x$, and so, contains either $2$ holes or none at all. These two cases give rise to two sets of feasible patterns as follows.
\begin {itemize}[label={--}]
\setlength \itemsep{1.5 mm}
 \item Case 1: $B$ contains $2$ holes of $S1$:\\
\noindent It follows that $B$ intersects with $5$ lines of $S1$ including $a_9$, and the possible intersection patterns are:\\

\textbf{1.} $ (3,\, 3, \, 1, \, 1) $; \hspace{3.6 mm} \textbf{2.} $ (2, \, 2, \, 2, \, 2) $; \hspace{3.6 mm} \textbf{3.} $ (3, \, 2, \, 2, \, 1) $.\\

As before, the above patterns are depicted as follows, with the line $a_9$ as the third column entry in each row (denoting a regulus).\\
\begin{center}
Pattern \textbf{1.}$\,\,\,
\begin{matrix}
\bullet &\bullet &\bullet\\ \bullet &\bullet &\bullet\\ \circ & \circ & \bullet \\ \circ & \circ & \bullet
\end{matrix} $ $ \,\,\,\,\,$ Pattern \textbf{2.} $\,\,\, \begin{matrix}
\circ &\bullet &\bullet\\ \circ &\bullet &\bullet\\ \circ &\bullet &\bullet \\ \circ &\bullet &\bullet
\end{matrix} 
$ $ \,\,\,\,\,$ Pattern \textbf{3.} $\,\,\, \begin{matrix}
\bullet &\bullet &\bullet\\ \circ &\bullet &\bullet\\ \circ &\bullet &\bullet \\ \circ &\circ &\bullet
\end{matrix}$\\
\end{center}
\item Case 2: $B$ does not contain any hole of $S1$\\
\noindent In this case, $B$ intersects with $7$ lines of $S1$, including $a_9$. The feasible intersection patterns are:\\

\textbf{1.} $ (3,\, 3, \, 2, \, 2) $; \hspace{4.8 mm} \textbf{2.} $ (3, \, 3, \, 3, \, 1) $.\\

The patterns are depicted as follows.\\
\begin{center}
Pattern \textbf{1.} $\,\,\,
\begin{matrix}
\bullet &\bullet &\bullet\\ \bullet &\bullet &\bullet\\ \circ &\bullet &\bullet \\ \circ &\bullet &\bullet
\end{matrix} $ $ \,\,\,\,\,$ Pattern \textbf{2.} $\,\,\, \begin{matrix}
\bullet &\bullet &\bullet\\ \bullet &\bullet &\bullet\\ \bullet &\bullet &\bullet \\  \circ &\circ &\bullet
\end{matrix}
$\\
\end{center}
\end{itemize}
\subsubsection{Elimination of Unrealizable Patterns}
The above intersection patterns were obtained solely on the possible inclusion of holes of one spread in a plane of the other spread. It turns out that all the patterns with the disjoint condition are realizable, as validated by examples in the next section. When a plane intersects with the common line, we prove that the third pattern of Case 1 cannot occur.

\bprop
Let $B$, a plane codeword, intersect with the common line $a_9$ of the spread $S1$, and contain two holes of $S1$. Then it cannot simultaneously intersect with all three lines of one regulus of $S1$, with two lines each of two others and with only $a_9$ of the fourth. 
\eprop
\begin{proof}
Let $B$ be explicitly described as the set $\{ x, y_1, y_2, y_3, y_4 , h_1, h_2 \}$, where $x \in a_9$, $y_i \in a_i$ for $i = 1,2,3,4$, and $h_1, h_2$ are holes of $S1$. Moreover, assume that $a_1, a_2$ belong to the regulus $r_1$ spanning the solid $\sigma_1$, while $a_3, a_4$ belong to different reguli $r_2, r_3$, with corresponding solids $\sigma_2, \sigma_3$, respectively. As $B$ contains holes of $S1$, it intersects with $\sigma_i, i = 1,2,3$, in lines $L_i, i = 1,2,3$, say. Hence these lines can be explicitly written as:\\
$L_1 = \{ x, y_1, y_2\}$, $L_2 = \{ x, y_3,x+ y_3\}$, $L_3 = \{ x, y_4,x+ y_4\}$.\\
As these lines are contained in the solids spanned by the reguli, the points $x+ y_3,x+ y_4$ cannot be identified with the holes $h_1, h_2$. If we assume that $x+ y_3 = y_1$ and $x+ y_4 = y_2$, we have: $y_3 = y_1 + x = y_1 + y_1 + y_2 = y_2$, a contradiction, and a similar conclusion follows from the other equation. Interchanging the assignments also leads to the same situation. Therefore, the lines $ L_2 , L_3$ cannot be defined as above and hence we conclude that this pattern cannot occur.
\end{proof}
We summarize the preceding results in the following theorem.
\bthm\label{th:intpat}
Let $\Cc$ be an optimal binary $(5,3)$ projective space code of size $18$, realized as $\Cc = S1 \cup (S2)^{\perp}$, where $S1, S2$ are maximal partial spreads of lines in $\pgft$. Further, let $S1$ be of type X with $a_9$ denoting the line common to all the reguli. Then a codeword plane $B \in (S2)^{\perp}$ exhibits any one of the following intersection patterns with the reguli of $S1$:
\begin{itemize}[label={--}]
	\setlength \itemsep{1.5 mm}
	\item $B$ is disjoint from $a_9$:
	\begin{enumerate}
		\item $B$ intersects with $2$ lines each of three reguli, and is disjoint from all the lines of the fourth regulus;
		\item $B$ intersects with $2$ lines each of two reguli and with one line each of the other two;
	\end{enumerate}
    \item $B$ meets $a_9$ in a point:
    \begin{enumerate}
    	\item $B$ intersects with all $3$ lines of two reguli and with only $a_9$ in the other two;
    	\item $B$ intersects with $2$ lines of every regulus;
    	\item $B$ intersects with all $3$ lines of two reguli and with $2$ lines, including $a_9$, in each of the other two.
    	\item $B$ intersects with all $3$ lines of three reguli and with only $a_9$ of the fourth regulus.
    \end{enumerate}	
\end{itemize}

Moreover, in the first two cases $B$ contains exactly $1$ hole of the spread $S1$, in the third and fourth, exactly $2$ holes, and in the final two cases, none at all.
\ethm
\begin{Remark}
{The preceding analysis does not use any particular property of the plane $B$ other than the fact that it is a plane codeword of an optimal code realized by doubling construction. Hence the possible intersection patterns apply to any of the nine codeword planes. However, as we show in Section \ref{sec:hokkpat}, specific constructions of optimal codes may impose additional restrictions on the intersection patterns exhibited by the codeword planes. }
\end{Remark}
\subsection{Examples of Patterns in Type X Optimal Codes}\label{sec:eggu}
We now present examples of optimal $(5,3)_2$ codes, formed by doubling construction using two type X spreads, such that they exhibit the intersection patterns described in the previous section. We have chosen examples using the particular dual plane $ B_9 = b_{9}^{\perp} $ of the common line of the second spread.\\
To obtain the codes, we created a database of maximal partial line spreads in $ \pgft $ as follows. First we constructed a representation graph, where each node corresponds to a unique line in $\pgft$, unique upto a representation as a $2 \times 5$ RREF matrix over $\bF_2$. An edge between two nodes guarantees no intersection between corresponding nodes. Starting from some seed type X maximal spread, other maximal line spreads were obtained via clique search on this graph, and those of type X chosen. Finally optimal doubling codes were obtained by pruned search among pairs of type X spreads.\\
\emph{Notation:} As used in \cite{RS}, we adopt the following compact representation for vectors in $\bF_{2}^{5}$. The canonical basis vectors are denoted by the indices $i = 1,2,\cdots, 5 $ for the single non-zero coordinates, along with the vector $11111$, denoted by $u$. An arbitrary vector is denoted by a tuple of its generating vectors. For instance, the vector $01001$ is denoted by the tuple $25$, and the tuple $3u$ stands for $11011$.\\
The spreads $ S1 , S2 $ consist of lines $ \{a_1, a_2, \cdots, a_9\} $
, $ \{b_1, b_2, \cdots, b_9\} $; the set of dual planes of $ S2 $ are denoted by: $ \{B_1, B_2, \cdots, B_9\} $. The regulus of $ S1 $ formed by the lines $ a_i, a_j, a_k $ is denoted by $ R_{ijk} $.
\subsubsection{Disjoint Patterns}
First we give examples of optimal codes such that the dual codeword $B_9$ of the second spread $S2$ is disjoint from the common line $a_9$ of the first spread $S1$.
\beg\label{egar1}
$S1 = \{1,25,125 \}, \{15,24,3u \}, \{14,23,5u \},\{145,234, 4u \},  \{12,$ \\ $345, u \}, \{124,34, 123 \}, \{2,35,235 \}, \{245,3,1u \}, \{135,4,2u \}$.\\
$S2 = \{1,23,123 \}, \{15,25,12 \}, \{14,1u,4u \}, \{145, 2, 3u\}, \{125,34,u \}, \{124,3,5u \}, $\\$ \{24,345,235 \}, \{245,35,234 \}, \{135,45,134 \}.$\\ 
Both these spreads are of type X, with $a_9$ and $b_9$ common to all $4$ reguli. The reguli of $S1$ are given by: $r_1 = R_{139}$, $r_2 = R_{249} $, $r_3 = R_{579}$ and $r_4 = R_{689}$. The dual plane $B_9 = (b_9)^{\perp}$ is disjoint from $a_9$ and all other lines of the regulus $r_1$. It intersects with the remaining two lines in each of the other reguli. 
\eeg
\beg
$S1 = \{1,25,125 \}, \{14,23,5u \},\{145,245,12 \}, \{135,1u,124 \}, \{13,$ \\ $ 45,2u \},  \{4u,4,$ $u \},  \{2,34,234 \}, \{3, 5, 35 \}, \{15,24,3u \}$.\\
$S2 = \{1,234,5u \}, \{15,2,125 \}, \{14,1u,4u \}, \{134,235,3u \}, \{124,34,123 \},  \{135, 4,$ $2u \},  \{25,45,24 \}, \{3, 5, 35 \}, \{145, 245, 12\}.$\\ 
As before, $a_9$ and $b_9$ are common to all $4$ reguli in the respective spreads. The reguli of $S1$ are: $r_1 = R_{139}$, $r_2 = R_{259} $, $r_3 = R_{679}$ and $r_4 = R_{489}$. The dual plane $B_9 = (b_9)^{\perp}$ is disjoint from $a_9$ and intersects with the remaining two lines in each of reguli $r_2 , \, r_4$. With respect to the reguli $r_1, \, r_3$, it intersects only with the lines $a_1$ and $a_6$, respectively.
\eeg
\subsubsection{Patterns with Intersection}
Now we give examples of optimal codes exhibiting the regulus-dual plane intersection patterns which arise when the dual plane codeword $B_9$ intersects with the common line $a_9$ of the first spread $S1$. 
\beg
Consider the pair of maximal partial spreads with the first spread $S1$ as in Example \ref{egar1} and the second spread given as follows.\\
$S2 = \{1,23,123 \}, \{15,25,12 \}, \{14,1u,4u \}, \{145, 2, 3u\}, \{125,34,u \}, \{124,3,5u \}, $\\$ \{24,345,235 \}, \{245,35,234 \}, \{13,45,2u \}.$\\ 
The second spread is again of type X, with $b_9$ common to all $4$ reguli. The dual plane $B_9 = (b_9)^{\perp}$ intersects with $a_9$ and all the remaining lines in each of reguli $r_3 , \, r_4$ of $ S1 $. With respect to the reguli $r_1, \, r_2$, it intersects only with $a_9$.
\eeg
\beg
Consider $S1$ as in Example \ref{egar1} and the second spread \\
$S2 = \{1,23,123 \}, \{15,25,12 \}, \{14,1u,4u \}, \{145, 2, 3u\}, \{125,34,u \}, \{124,3,5u \}, $\\$ \{24,345,235 \}, \{245,35,234 \}, \{13,5,135 \}.$\\ 
$ S2 $ has the line $b_9$ common to all $4$ reguli. The dual plane $B_9 = (b_9)^{\perp}$ intersects with $2$ lines, including $a_9$, in each of the $4$ reguli. It is disjoint from the lines  $a_1, a_4, a_5, a_8$ of the reguli $r_1,  r_2, r_3, r_4$, respectively.
\eeg
\beg
Consider the following pair of maximal partial spreads.\\
$S1 = \{1,25,125 \}, \{15,24,3u \}, \{14,23,5u \},\{145,234,4u \}, \{12,345,u \}, \{124,34,$ $123 \}, \{2,35,235 \}, \{245,3,1u \}, \{13,5,135 \}$.\\
$S2 = \{1,23,123 \}, \{15,25,12 \}, \{14,1u,4u \}, \{2u, 235, 124\}, \{145,35,134 \}, \{125,4,$ $ 3u \},  \{2,34,234 \}, \{3,45,345 \}, \{135,245,5u \}.$\\ 
Both these spreads are of type X, with $a_9$ and $b_9$, respectively, common to all $4$ reguli. The regulus structure of $S1$ is given by: $r_1 = R_{179}$, $r_2 = R_{289} $, $r_3 = R_{359}$ and $r_4 = R_{469}$. The dual plane $B_9 = (b_9)^{\perp}$ intersects with all the lines of the reguli $r_1, r_3$ and with two lines each of the reguli $r_2, r_4$, being disjoint from the lines $a_6, a_8$, respectively.
\eeg
\begin{Remark} 
We have not, so far, obtained any instantiation of the last pattern of Theorem \ref{th:intpat}. But we have not been able to theoretically rule out this particular pattern either.
\end{Remark}
\section{Characterization of the HKK Construction}\label{sec:hokkpat}
In this section we determine the intersection patterns exhibited by the dual codeword planes in the HKK construction \cite{HKK} for optimal binary $(5,3)$ projective space codes.\\

An HKK optimal code consists of the set obtained by \emph{point-hyperplane shortening} of a $(6,3,2)$ lifted Gabidulin code augmented by two planes in $PG(5,2)$ which are at subspace distance $4$. One of these extra planes, $E' = E_2$ in the notation of Honold \emph{et al.}, belongs to the shortening hyperplane $H = \pgft$, and hence, becomes a codeword plane. Hence $E'$ is the ninth plane of the optimal code, which is $B_9$ in our notation. The other plane, $E $ in the HKK notation contains the shortening point $P \notin H$. On shortening, it produces the ninth line $L_1 = E \cap H$, which we label $a_9$. The line $L_1$ is the `moving line' of the `special plane' $E_1 = (\Sc +P)\cap H$, which contains the holes of the spread $S1$, $\Sc$ being the plane disjoint from all codewords of the lifted Gabidulin code in $PG(5,2)$. In our notation, $E_1 = \alpha_9$.\\
From the details of the construction, it is surmised that the eight dual plane codewords and the ninth plane $B_9$ may exhibit different intersection patterns with the lines of the first spread $S1$. The following propositions show that such is indeed the case.
\bprop
In the HKK construction, restricted to optimal binary $(5,3)$ codes, the dual $B_9 \in (S2)^{\perp}$ meets $ a_9 \in S1 $ in a point, and can exhibit only the following intersection patterns.\\
\textbf{1.} $ (3,\, 3, \, 1, \, 1) $; \hspace{4.8 mm} \textbf{2.} $ (2, \, 2, \, 2, \, 2) $.\\

\eprop
\begin{proof}
From the description of the HKK construction, the additional plane in the optimal code is chosen to contain the line $\Sc \cap H$ and the additional line $ a_9 = L_1 $ is such that it is contained in $(\Sc +P)\cap H$. It follows that $B_9 \cap \Sc = \Sc \cap H = L_2$, say, and the special plane $ E_1 = \alpha_9 =  a_9 + L_2$, as $L_2 = \Sc \cap H \subset (\Sc +P)\cap H $. Therefore, $a_9 $ and $ L_2 $ meet in a point which is also identified with $B_9 \cap a_9$.\\		
Therefore, in the HKK construction, the codeword plane $B_9$ intersects with the common line $a_9$ and shares a line $L_2$ with $ \alpha_9$, thereby containing two holes of the spread $S1$. The assertion then follows from Theorem \ref{th:intpat}.
\end{proof}
\bprop
In the HKK construction for optimal $(5,3)_2$ codes, the plane codewords $B_i, i = 1,2,\cdots, 8$, can exhibit the following intersection patterns.
%
%
%
\begin{enumerate}[label = (\roman*)]
	\setlength \itemsep{1.5 mm}
	\item $B$ is disjoint from $a_9$:\\	
	\textbf{1.} $ (2,\, 2, \, 2, \, 0) $; \hspace{4.8 mm} \textbf{2.} $ (2, \, 2, \, 1, \, 1) $; 
	\item $B$ meets $a_9$ in a point:\\	
	\textbf{3.} $ (3, \, 3, \, 2, \, 2) $.; \hspace{4.8 mm} \textbf{4.} $ (3, \, 3, \, 3, \, 1) $.
	
\end{enumerate}

\eprop
\begin{proof}
Each one of the eight remaining codeword planes $B_i, i = 1, \cdots, 8$, is a codeword of the lifted Gabidulin code contained in $H$. Hence they are all disjoint from $\Sc$ and consequently, from the line $L_2 = \Sc \cap H $. Thus they intersect with the special plane $\alpha_9$  of the spread $S1$ in a \emph{single point} of $\alpha_9 \setminus L_2$. This point can be one of the two holes of $S1$ outside $L_2$ or it can be one of the two points of $L_1 = a_9$, other than the point $a_9 \cap L_2$.\\
Therefore, in the HKK construction, each plane codeword $B_i, i = 1, \cdots, 8$, is either disjoint from $a_9$ and intersects $\alpha_9$ in a single hole of the spread $S1$, or it intersects with the common line $a_9$ and does not contain any hole at all. The assertion now follows from Theorem \ref{th:intpat}.
\end{proof}
This completes the description of the possible intersection patterns of the HKK construction.
\section{Conclusion}
The largest class of optimal binary $(5,3)$ subspace codes may be interpreted in terms of the doubling construction: $S1 \cup (S2)^{\perp}$, where $S1, S2$ are two maximal partial line spreads in $\pgft$. We have first identified the types of the line spreads in the doubling instances of optimal $ (5,3)_2 $ codes obtained from two recent constructions: one by Honold \emph{et al.} (HKK) and the other by Cossidente \emph{et al.} (CPS). Next we have presented a novel characterization of doubling codes when both the spreads are of type X. Our classification is based on the intersection patterns among the reguli of the spread $S1$ and the codeword planes which are the duals of the lines of the spread $S2$. As an application, we have obtained the realizable patterns in the instances of doubling in the HKK construction for these parameters. A similar study on the doubling codes obtained from the CPS construction, i.e., the case of the $\mathrm{I}\Delta- \mathrm{I}\Delta $ doubling codes, is underway.\\

\noindent As has been stated before, compared to the other existing techniques for constructing optimal $ (5,3)_2 $ subspace codes, the doubling construction has the decided advantage of simplicity. Given a repository of all possible maximal line spreads, it boils down to simply checking for the `compatible' pairs that make up an optimal code via doubling. The present article has partially succeeded in providing a set of conditions which will result in obtaining such pairs.\\
It has been proved in the earlier sections that the doubling instances obtained from the two existing constructions generally result in either Type X or Type $ \mathrm{I}\Delta $ codes. Our study of doubling codes, obtained using a search on maximal spread pairs, yields the following observation. Optimal codes tend to occur in clusters with minimal changes resulting in completely different reguli structures and hence, different constituent spread types. The results presented in this article thus form the first step towards characterizing the intersection patterns among the reguli and lines of all types of constituent spreads. This will lead to a complete set of conditions to determine when ``doubling'' two arbitrary MLS's in $ \pgft $ will yield an optimal binary $ (5,3) $ code.
%
\bibliography{bibin}
 
\end{document}